\documentclass[prl,aps,twocolumn,amssymb,showpacs]{revtex4}
\usepackage{amsmath}
\usepackage{graphicx}
\usepackage{graphics}
\usepackage{epsfig}
\usepackage{amsfonts}
\usepackage{amssymb}
\usepackage{latexsym}

\begin{document}

\pacs{05.45.-a; 05.60.Cd}
\title{Normal and anomalous diffusion in random potential landscapes}
\author{F. Camboni and I.M. Sokolov}
\affiliation{Institut f\"ur Physik, Humboldt-Universit\"at zu Berlin, Newtonstr. 15, D-12489 Berlin, Germany}

\begin{abstract}
A relation between the effective diffusion coefficient in a lattice with random site energies and random trasition rates 
and the macroscopic conductivity in a random resistor network allows for elucidating possible sources of anomalous diffusion 
in random potential models. We show that subdiffusion is only possible either
if the mean Boltzmann factor in the corresponding potential diverges or if the percolation concentration 
in the system is equal to unity (or both), and that superdiffusion is impossible in our system under any condition.
We show also other useful applications of this relation.  
\end{abstract}

\maketitle

A classical particle's diffusion in a random potential or hopping on a lattice with
disordered site energies $E_i$ is a versatile theoretical model with a wide range of applications \cite{Bouchaud}. 
The particle's motion typically corresponds to normal diffusion, but can get subdiffusive 
in the presence of deep traps or in the case of infinte contrast when
approaching percolation transition. A question arises, whether there can be other cases leading
to subdiffusion except for these two (or combinations thereof).
In what follows we show that the diffusion coefficient in a discrete disordered lattice is 
always finite (i.e. that no superdiffusion can be observed) 
but may vanish (possibly giving rise to subdiffusion).
Independently of the particular distribution of \textit{nonzero} transition rates, 
this is possible either if the percolation threshold in the corresponding network is unity 
(e.g. in one dimension, on finitely ramified fractals, or when we are \textit{already at percolation threshold}) 
or if the mean Boltzmann factor $\langle \exp(-E_i/kT) \rangle$ diverges (or both). 
According to the Arrhenius law, the last situation corresponds to the divergence 
of the mean sojourn time at a site, pertinent to trapping.

We start from the Master equation 
for the probabilities $p_i$ to find a particle at a site $i$ of a lattice (with lattice spacing $a$)
\begin{equation}
\dot{p}_i = \sum_j \left(w_{ij}p_j - w_{ji} p_i \right),
\label{Meq0}
\end{equation}
where $w_{ij}$ are transition rates from site $j$ to site $i$ different from zero
only for nearest neighbors. Eq. (\ref{Meq0}) can either 
follow from some microscopic scheme or be obtained by a discretization of the Fokker-Planck equation 
for the overdamped motion in a continuous potential. The system is taken to be homogeneous and isotropic in 
statistical sense. This requirement
excludes underdamped cases for which the velocities and coordinates enter differently, thus 
leading to anisotropy of the state (phase) space. In what follows we consider a $d$-dimensional lattice with 
total of $M \gg 1$ sites assigned energies $E_i$ being identically distributed random variables. 
We assume that the system is isothermic and possesses true thermodynamical equilibrium under appropriate boundary conditions, 
i.e. that the transition rates fulfill the detailed balance condition $w_{ij} p_j^0 = w_{ji} p_i^0$ 
at equilibrium (the superscript 0 will denote the corresponding value at equilibrium throughout the work).
The transition rates $0 \leq w_{ij} < \infty$ are not necessarily bounded from above,
and some of them may be put to zero to mimic percolation situations.
The values of $p_i^0$ are given by $p_i^0 \propto b_i= \exp(-E_i/kT)$, where $b_i$ denotes the Boltzmann factor, 
$T$ is the temperature and $k$ is the Boltzmann constant. 

Our discussion maps the initial problem onto the one for random resistor-capacitor networks. 
Let $g_{ij}$ be the corresponding conductivities of the bonds, and 
$\left\langle g_{ij} \right\rangle_{EM}$ be the effective conductivity
of a corresponding network in the static regime. 
Then the effective diffusion coefficient in a network follows as  
\begin{equation}
D^* = a^2 \frac{\left\langle w_{ji} \exp(-E_i/kT) \right\rangle_{EM}}{ \left\langle \exp(-E_i/kT)\right\rangle}. 
\label{MainEq}
\end{equation}
The statements done in the first paragraph are then demonstrated 
by using the results from the theory of electric circuits and from the percolation theory. 
Some other useful applications of Eq.(\ref{MainEq}) are shown.  

Eq.(\ref{MainEq}) by itself is not new, but we give here its physical derivation 
which stresses its general applicability and its connection with thermodynamics.
Thus, the discussion for the case of a barrier model (all $E_i$ are the same, but the transition rates
fluctuate) is contained in Ref. \cite{Bouchaud}, Eq.(2.15). 
Moreover, Eq.(\ref{MainEq}) naturally appears when applying effective medium approximation (EMA), like the one of \cite{Karayiannis}. 
A derivation for a continuous case (Langevin description in Ito interpretation) is given in \cite{Dean}. 
Note that the Ito prescription may correspond to the trap model in the discrete case \cite{Sokolov}, i.e. to a situation
different from the one of Ref. \cite{Bouchaud}. In our work we confine ourselves to a discrete setup
which allows for the application of the theory of electric circuits for the analysis of the results
(although we make a limiting transition to continuum to illustrate some outcomes of the approach).   

We start by rewriting Eq.(\ref{Meq0}) as 
an equation for mean numbers (``concentrations'') of non-interacting particles at the corresponding sites, 
$\dot{n}_i = \sum_j \left(w_{ij}n_j - w_{ji} n_i \right)$,
connected with probabilities via $n_i = N p_i$ with $N$ being the total number of particles.
In equilibrium all $n_i^0$ are proportional to the Boltzmann factors, $n_i^0 =C b_i$ with prefactor 
$C$ depending on the number of particles, on the system's size and on distribution of $b_i$.
Putting the detailed balance condition into the form $w_{ij} n_j^0 = w_{ji} n_i^0$ we denote
$w_{ij} n_j^0 = g_{ij}$ where $g_{ij}=g_{ji}$ is now a property of the bond. 
Using this notation we rewrite Eq.(\ref{Meq0}) as an equation for the temporal evolution of 
\textit{activities} $\zeta_i = n_i/n_i^0$ (see the Appendix): 
\begin{equation}
\dot{\zeta}_i =  \frac{1}{n_i^0} \sum_j  \left( g_{ij} \zeta_j - g_{ji} \zeta_i \right).
\label{Fuga}
\end{equation}
Eq.(\ref{Fuga}) is formally equivalent to the evolution equation of node potentials in a 
random resistor-capacitor model \cite{Bouchaud}, with
conductivities $g_{ij}$ and capacitances $n_i^0$.

Let us now calculate the effective diffusion coefficient provided it exists
(i.e. the system homogenizes at large scales).
For random resistor networks the homogenization of conductivity is mathematically proved for local conductivities bounded
from above and from below, see \cite{Chayes} and references therein. The boundness from below excludes 
the conductor-isolator percolation model, but homogenization still holds \textit{provided} the system percolates \cite{Mathieu}.
Physically, it is known that the conductor-superconductor system homogenizes below the percolation threshold for
superconductor \cite{Wright}.    

We mimic a stationary experiment on measuring the diffusion coefficient via the first Fick's law:
The system, in a form of a membrane of thickness $L$ 
and of transversal dimension $W$ separates two reservoirs, 
the left one with a well-stirred solution of particles 
at concentration $\nu_l$, the other one with a slightly lower concentration $\nu_r$.
The membrane is considered as impermeable for the solvent, and both concentrations are kept constant. 
The constant particles' current $I$ through the membrane is measured and connected with the mean diffusion coefficient inside it. 
Since in general a jump of the (free) energy per particle can form on a contact between the
membrane and the solution (e.g. when the fluid is a good solvent for diffusing particles and the membrane is, 
on the average, a bad one, or other way around) the effective 
diffusion coefficient \textit{inside} the membrane has to be defined through
\begin{equation}
D^* = \frac{I L}{W^{d-1} (\langle n_l \rangle - \langle n_r \rangle)}
\label{Exp}
\end{equation}
where $\langle n_l \rangle$ and $\langle n_r \rangle$ are the mean particle concentrations in the layers of 
the membrane in immediate contact with the solution, see Fig.1. In a stationary state $\dot{\zeta_i}=0$. Moreover, 
in the thermodynamical limit $L \to \infty$ the permeability of the membrane and thus the current tend to zero. 
We will call this situation ``quasi-equilibrium'' in what follows.

\begin{figure}[h!] 
\begin{center}
\includegraphics[width=8.3cm]{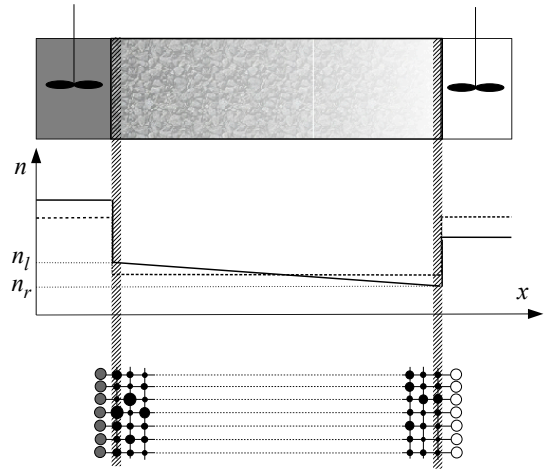}
\caption{A schematic illustration of the situation considered in the text: the disordered medium in contact with two
reservoirs, the mean concentration at different positions and the lattice model applied.}
\label{Fig}
\end{center}
\end{figure}

The contact with solution is modelled by additional arrays of sites to the left and to the right from the
membrane, with constant particles' concentrations and constant energies $E_0$ which can be chosen
arbitrarily ($E_0$ defines the quality of the solvent).
These additional sites are connected to the ones on the membrane's sides via extremely high 
transition rates fulfilling the detailed-balance condition.
In this case a local equilibrium between the surface sites and the solutions 
persists independently on the particles' distribution inside the bulk. Due to this
the activities of the surface sites are all equal to
$\zeta_l = A \nu_l$ and $\zeta_r = A \nu_r$ at the left resp. right boundary 
of the membrane, where the prefactor $A$ depends on $E_0$. Thus, $n_i$ in the leftmost
layer are proportional to $\zeta_l n_i^0 = A \nu_l n_i^0$ and in the rightmost layer to $\zeta_r n_i^0 = A \nu_r n_i^0$ so that 
the mean concentrations in the layers are $n_l =  A \nu_l \langle n_i^0 \rangle$ and $n_r =  A \nu_r \langle n_i^0 \rangle$,
where we assume that the distribution of the site energies in the surface layers is the same as in the bulk. 
We then calculate the corresponding total current $I$ through the system noting that the
equations for the currents and activities in a stationary state are the same as the ones given by the 
Kirchhoff's laws for an electric circuit. Making such a reinterpretation we see that
$I = g^* (W^{d-1}/L) (\zeta_l - \zeta_r)$
where $g^*=\langle g \rangle_{EM}$ is the effective conductance (conductivity of a bond in the 
\textit{effective ordered medium} with the same total conductivity as our heterogeneous one), where the subscript EM denotes the effective medium mean.
Therefore $D^* =  a^2 g^* / \langle n_i^0 \rangle$. The prefactor $a^2$ is introduced to restore 
the dimension as follows from Eq. (\ref{Exp}), when passing from distances $L$ 
measured in lattice units to distances measured in centimeters. 
We note that since $n_i^0$ are proportional to the Boltzmann factors,
and since rescaling of all $g_{ij}$ by a constant factor leads to changing $g^*$ by the same factor, the 
proportionality factor $C$ cancels out;  this gives Eq.(\ref{MainEq}). 

The result is rather transparent. If we are able to measure
the effective conductivity of the system, we can connect it with the effective mobility $\mu^*$ 
(and thus with the diffusion coefficient) via Nernts-Einstein equation $\sigma^* = n_0 q \mu^* = n_0 q D^*/kT$, 
where $n_0$ is the equilibrium concentration of particles with charge $q$. 
Reverting this expression we get $D^* \propto \sigma^* / n_0$, which is essentially Eq.(\ref{MainEq}). 

One may argue that the correct way is to define $D^*$ through
the gradient of the coarse-grained concentration, and not via the total concentration difference.
As we show in the Appendix, this definition leads to the same result since local concentrations
and local activities decouple under quasi-equilibrium (but \textit{only} under this condition!). 

Let us first discuss some applications of Eq.(\ref{MainEq}) other than discussed in the preface.  
Eq.(\ref{MainEq}) gives the possibility to obtain the universal
bounds on the effective diffusion coefficient based on those for
the effective conductance, i.e. the universal Wiener bounds \cite{Wiener} and
the tighter Hashin-Shtrikman bounds for isotropic systems
\cite{HashinShtrikman}, as well as to generalize some exact results for
two-dimensional systems based on duality \cite{Mendelson}.
In this presentation we concentrate on continuum models
where Eq.(\ref{Meq0}) arises from discretization of a Fokker-Planck equation
for $p(\mathbf{x},t)$: $\dot{p} = D \Delta p + (D/kT)\nabla (\nabla U(\mathbf{x})p)$
with constant diffusion coefficient $D$ and disordered potential $U(\mathbf{x})$.
The details of calculations are given in the Appendix. 

The universal Wiener bounds for the conductance are given by 
$\langle g_{ij}^{-1} \rangle^{-1} \leq \left\langle g_{ij} \right\rangle_{EM} \leq \langle g_{ij} \rangle$,
In our cases this corresponds to
\begin{equation}
\frac{a^2 w_0}{\langle \exp(E_i/kT) \rangle \langle \exp(- E_i/kT) \rangle} \leq D^* \leq  a^2 w_0.
\end{equation}
Note that the lower bound reproduces the exact result for the one-dimensional system
with random potential and constant diffusion coefficient.

\begin{figure}[h]
\begin{minipage}{4.0cm}
\begin{tabular}{c}
\includegraphics[height=4.0cm]{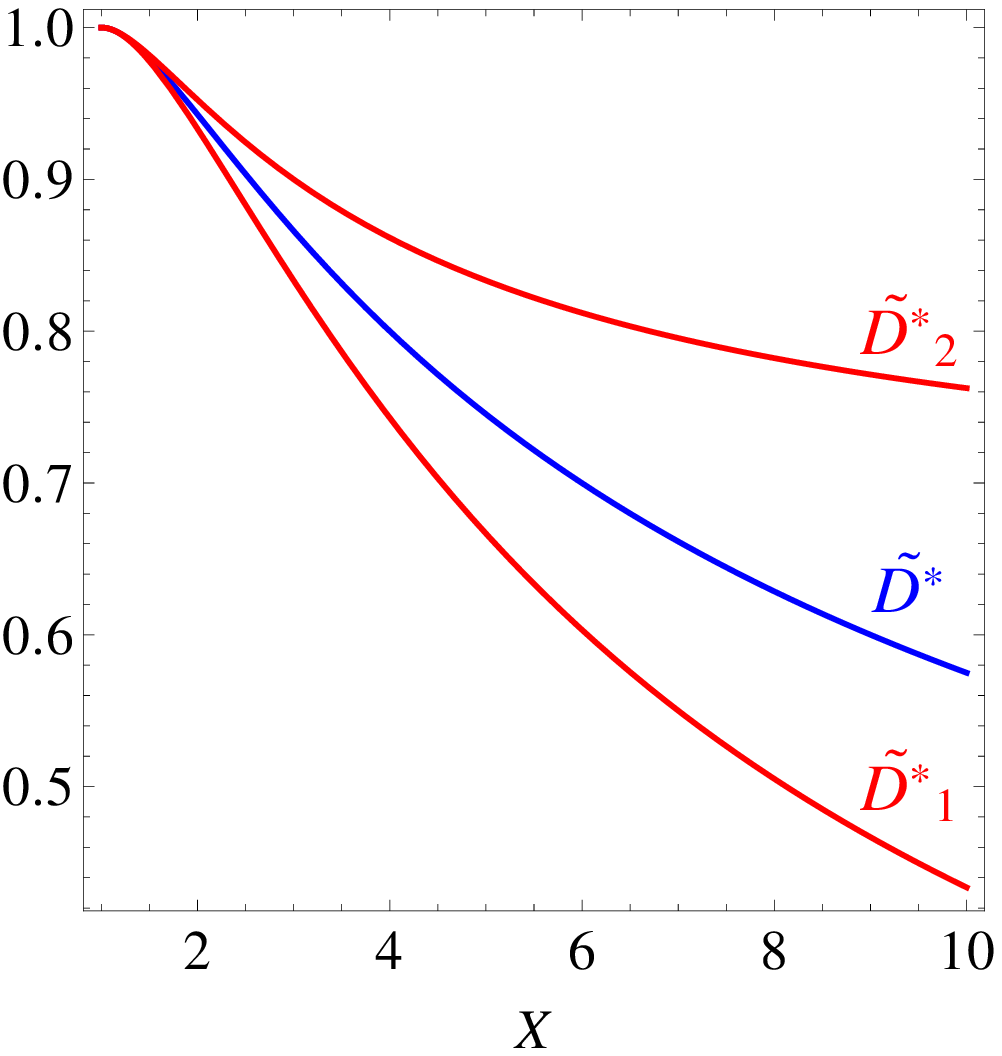} \\
(a)
\end{tabular}
\end{minipage}
\hfill
\begin{minipage}{4.0cm}
\begin{tabular}{c}
\includegraphics[height=4.0cm]{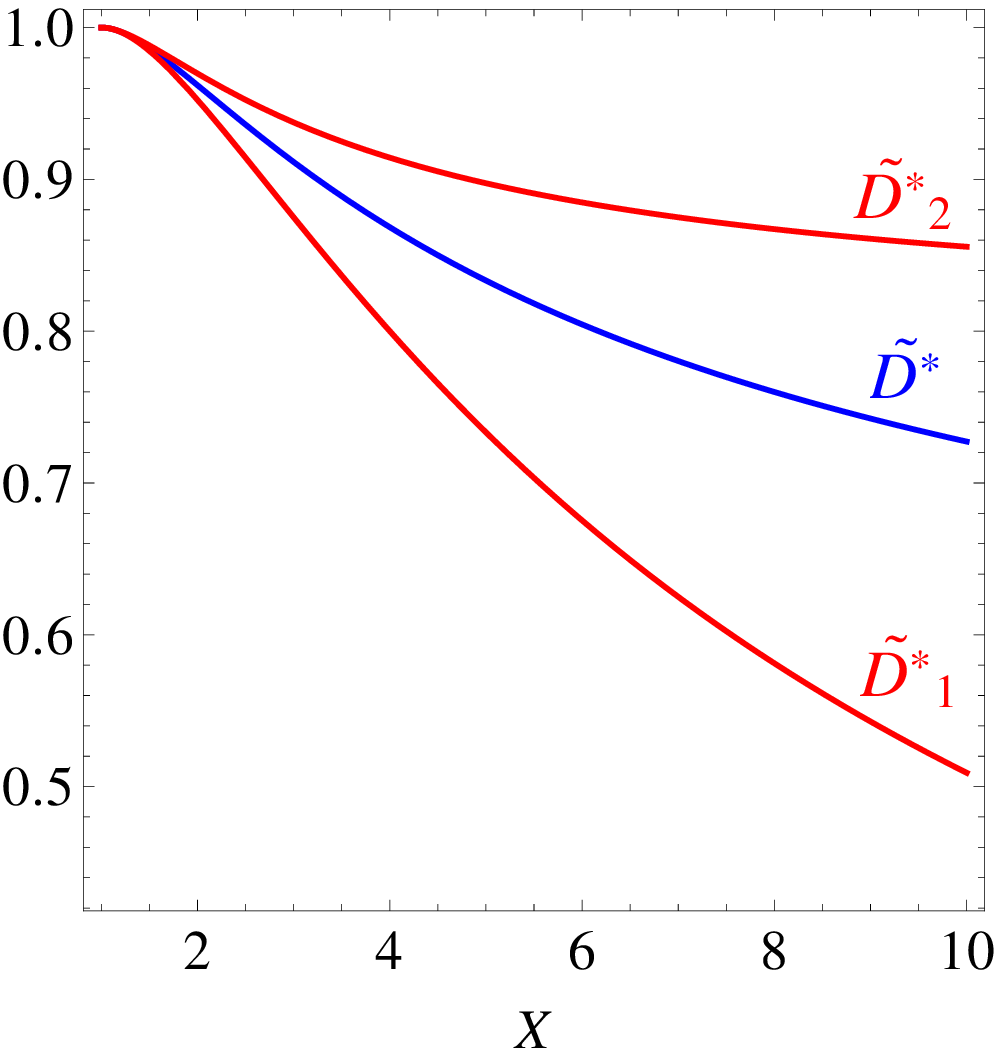} \\
(b)
\end{tabular}
\end{minipage}
\caption{EMA result and Hashin-Shtrikman bounds for the 
symmetric binary case ($p=1/2$) vs. contrast $x$ for: 
$d=2$ (a) and for $d=3$ (b). }
\label{Fig:BinHS}
\end{figure}

\begin{figure}[h]
\begin{minipage}{4.0cm}
\begin{tabular}{c}
\includegraphics[height=4.0cm]{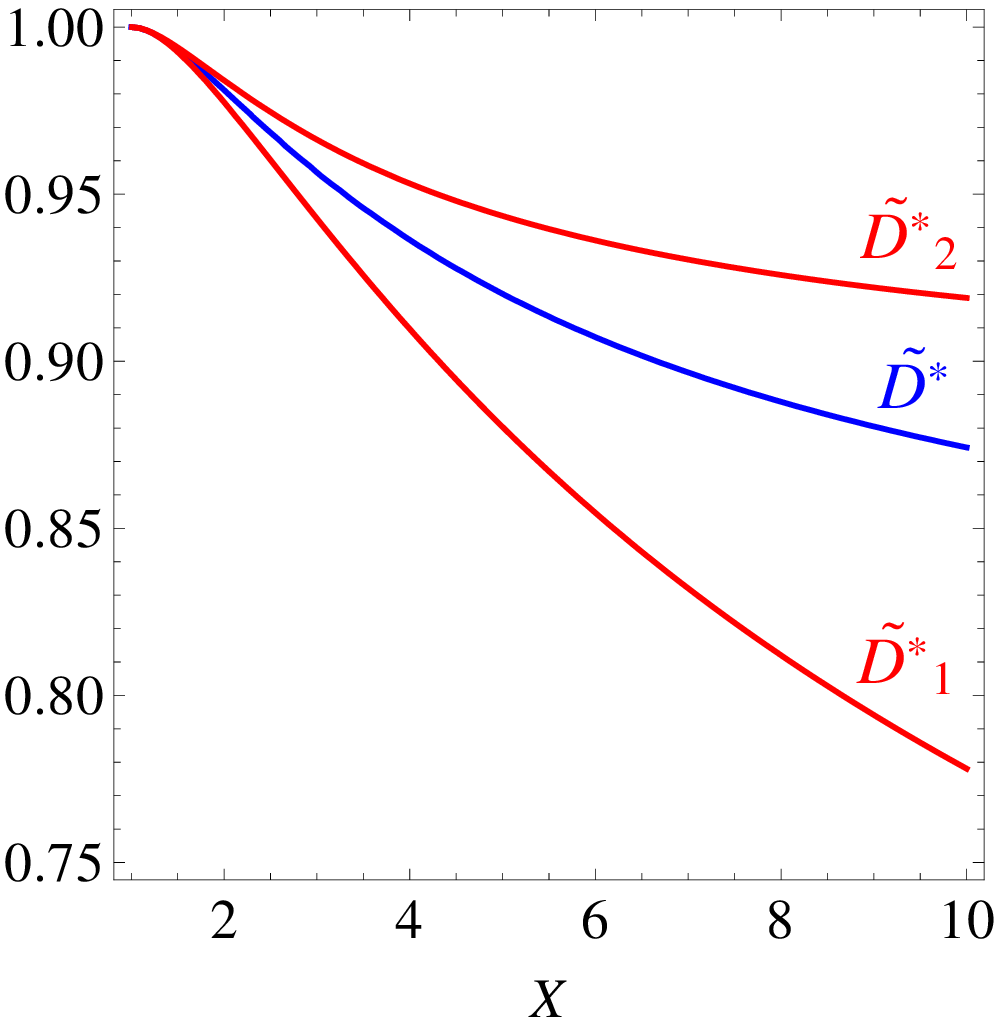} \\
(a)
\end{tabular}
\end{minipage}
\hfill
\begin{minipage}{4.0cm}
\begin{tabular}{c}
\includegraphics[height=4.0cm]{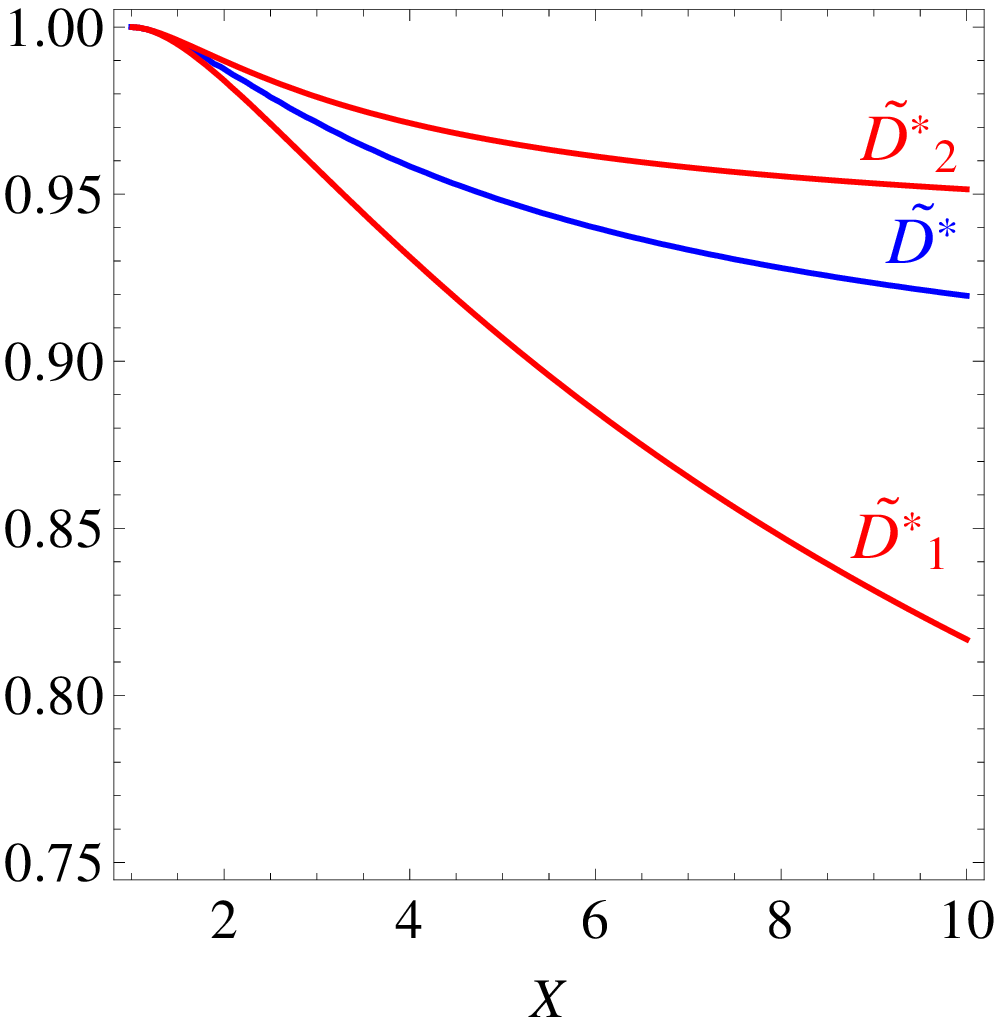} \\
(b)
\end{tabular}
\end{minipage}
\caption{EMA result and Hashin-Shtrikman bounds for
truncated exponential $P(E_{i})$ vs $x$ for $d=2$ (a) and $d=3$ (b).}
\label{Fig:HomoHS}
\end{figure}

In Figs. \ref{Fig:BinHS} and \ref{Fig:HomoHS} we plot the Hashin-Shtrikman  bounds for two cases: the case of the
binary disorder $E_i = E_1$ with probability $p$ and $E_i = E_2$ with probability $1-p$, 
and the case of $E_i$ possessing an exponential distribution with cutoffs, the one with density
$P(E_{i}) = \beta \, e^{-\beta E_{i}} [e^{-\beta E_2} - e^{-\beta E_1}]^{-1}$
for $E_2 < E_i < E_1$ and vanishing elsewhere (apart from cutoffs this distribution is reminiscent
of the exponential energy distributions leading to CTRWs).  As a comparison, the
results of the effective medium approximation (EMA), see Refs. \cite{HausKehr,Kirkpatrick}, are shown.
The results are plotted as the function of a contrast $x = \exp[(E_2-E_1)/kT]$
being the ratio of the maximal and the minimal value of $g_{ij}$. 

Note that the result for the effective diffusion coefficient for symmetric binary case in 2d
is essentially exact since in this case $\langle g \rangle_{EM} = \sqrt{g_a g_b}$ due
to the duality relation \cite{Mendelson}.

In the limit of very strong disorder $D^*$ may vanish or diverge. In the first case $D=0$, the system either does not show any
transport (does not percolate) or shows anomalous transport slower than diffusion (i.e. shows \textit{subdiffusiom}). In the second case
it might show superdiffusion.

If $D^*$ vanishes, it can do so either because the numerator  
$\left\langle w_{ji} \exp(-E_i/kT) \right\rangle_{EM}$ vanishes or because 
the denominator $\left\langle\exp(-E_i/kT)\right\rangle$ diverges, as well as in the cases when both possibilities 
are realized simultaneously (which may give rise to subdiffusion of mixed origins \cite{Meroz}). 
We discuss the conditions under which the corresponding behavior may take place. 
If $D^*$ diverges, it can do so because the numerator diverges, or because the denominator vanishes, or both. 
As we proceed to show, none of these possibilities can be realized.
This excludes not only superdiffusion, but also the ``compensated'' cases of normal diffusion when both
numerator and denominator vanish or diverge simultaneously. 

For further discussion we first recapitulate the following properties of percolation systems:
(i) The mixture of resistors with given finite conductivity (at concentration $p$) with insulating bonds (of zero conductivity)
at concentration $1-p$ possesses zero conductance below the percolation threshold $p_c$ and finite conductance above it. The corresponding 
system homogenizes at scales above the correlation length \cite{Bouchaud}. This homogenization also takes place for
arbitrary distribution of the conductivities of the resistors \cite{Mathieu}. Similarly,
(ii) The mixture of resistors with given finite conductivity (at concentration $p$) with superconducting bonds (at concentration $1-p$)
possesses finite conductance below the percolation threshold $p_c^s$ for superconducting bonds, with $1-p_c^s = p_c$, 
and infinite conductance above it.
These properties do hold not only for the Bernoulli percolation model but also in the case 
when the short-range correlations in the occupation probabilities
of the bond by the corresponding resistors / insulators / superconductors are present.
This statement is a (silently assumed) basis of all renormalization group approaches in percolation. 

Using the results of the theory of electric circuits, see e.g. \cite{Newstead}, we show that the 
total conductivity of a resistor network is a non-decaying function of the conductivity
of each particular bond. Let us consider the system as
placed between two ``superconducting'' bars considered as a terminal 1 of the system. Let us consider the poles
$i$ and $j$ between which $g_{ij}$ is switched as terminal 2.
Using the theory of two-terminal circuits we calculate the input impedance (total conductivity) $z_{\mathrm{in}}$
as a function $g_{ij}$: $z_{\mathrm{in}} = z_{11} - z_{12}z_{21}/(z_{22}+g_{ij})$,
where $z_{\alpha \beta}$ are the elements of the impedance matrix of the system. For a system of reciprocal passive
elements (no batteries, no diodes) this matrix is non-negatively definite and symmetric as a consequence of 
non-negative heat production and of reciprocity theorem. In the case of pure resistor network the matrix is real. Thus,
$z_{11},z_{22} \geq 0$ and $z_{12}z_{21} = (z_{12})^2 \geq 0$, 
so that $z_{\mathrm{in}}$ is a non-decaying function of $g_{ij}$. 

Now we show that the numerator never diverges. We fix some $q < 1-p_c^s = p_c$ and declare the fraction
$q$ of bonds (starting from the ones with largest $g$) to be superconductive. The lowest conductivity
of a changed bond is $g_{\mathrm{min}}$. The superconducting bonds are non-percolating by construction, 
and the conductance of the remaining system is finite, being smaller than a conductance of the resistor-superconductor
mixture where all conductivities are put to  $g_{\mathrm{min}}$. Thus the numerator can only diverge if
$p_c=0$, i.e. never in finite dimension.

The numerator does not vanish for a system with percolation
concentration $p_c<1$. Let us remove a portion  $q < 1- p_c$ of bonds 
with smallest $g$ without destroying percolation and denote the largest removed conductivity by $g_{\mathrm{max}} > 0$. 
The rest of the system percolates and has a conductance which is larger then the conductance of a two-phase
system constructed of resistors with  $g = g_{\mathrm{max}}$ and $g = 0$, which is nonzero since we are 
above percolation threshold. Thus the numerator can only vanish if $p_c=1$, and no bonds can be removed.

The denominator in Eq.(\ref{MainEq}) can diverge if the corresponding mean value of the Boltzmann factor diverges.
Since $b_i=\exp(-E_i/kT)$ is proportional to the sojourn time at a site $i$ in equilibrium, this corresponds to
diverging mean sojourn time at a site, i.e. to a trap model, which in high dimensions is equivalent to CTRW with a broad distribution of waiting times. 

The denominator can not vanish. Let $p(E)$ be the probability density of $E_i$, and
$E_M > -\infty$ its median, $\int_{-\infty}^{E_M} p(E)dE = 1/2$. Since
$\langle b \rangle = \left\langle\exp(-E_i/kT)\right\rangle = \int_{-\infty}^\infty e^{-E/kT} p(E)dE$ where the 
integrand is non-negative, we have $\langle b \rangle  
> \int_{-\infty}^{E_M}e^{-E/kT} p(E)dE > \exp(-E_M/kT) \int_{-\infty}^{E_M} p(E)dE =(1/2) \exp(-E_M/kT)$
since $\exp(-E/kT)$ is monotonically decaying.

Summarizing our findings we state that there exists an exact correspondence between the effective diffusion coefficient in a random 
potential and macroscopic conductivity in a random resistor model. This simple relation allows us to obtain exact
bounds on the effective diffusion coefficient. It also
allows for elucidating possible sources of anomalous diffusion in such model. Thus, the subdiffusion is possible either
if the mean Boltzmann factor of the corresponding potential diverges (energetic disorder) or if the percolation concentration 
in a system is equal to unity, i.e. if the system is already at the
percolation threshold, in one dimension, or on finitely ramified fractals (structural disorder), 
and that superdiffusion is impossible in our system under any condition.

\textbf{Acknowledgement.} The work was supported by BMU within the  
project "Zuverl\"assigkeit von PV Modulen II".

\bigskip

\appendix

\section{Appendix}

\subsection{A - Activities and chemical potentials}

The idea behind the approach is based on the fact that the probabilities $p_i$  depend on the system's
size $M$ and therefore are awkward to use in mimicking a macroscopic experiment. 
Thus, according to normalization condition, $\sum_{i=1}^M p_i =1$, and in equilibrium
$p_i^0 = b_i/\sum_{i=1}^M b_i$ or $p_i^0 = M^{-1} b_i / \langle b_i \rangle$ provided the
corresponding mean exists. On the other hand, $n_i = N p_i$ and $n_i^0$ are
the ``intensive'' variables, e.g. $n_i^0 = (N/M) b_i / \langle b_i \rangle$, and do not depend on the system's 
size provided the mean particle concentration $N/M$ is kept constant. 

To see that $\zeta_i$ introduced in the main text are indeed activities one can proceed as follows. 
Adding a weak external potential (e.g. electric field giving rise 
additional potential energy $V_j$ at site $j$) changes the equilibrium concentrations and hence the corresponding equation to 
$\dot{n}_i = \sum_j  \left[g_{ij}(n_j/n_j^0)\exp(V_j/kT) - g_{ji} (n_i/n_i^0) \exp(V_i /kT)\right]$.
Here $\zeta_i=(n_i/n_i^0) \exp(V_i /kT)$ can indeed be interpreted as activities, since 
$\mu_i = kT \ln \zeta_i = kT \ln \frac{n_i}{n_i^0} + V_i$
have a form of chemical potentials on the sites. 

The master equation can be interpreted as a combination of the local continuity equation
$\dot{n}_i = \sum_j J_{ij}$
and the local linear response equation $J_{ij} = g_{ij} (\zeta_j -\zeta_i)$ with $\zeta_j = n_j \exp(V_j/kT) / n_j^0$.
Close to equilibrium and for very small potentials $V$ the values of $\zeta$ are close to unity,
so that the deviations of chemical potentials from their equilibrium values of zero are small. Therefore, close to equilibrium,
\[
\dot{n}_i = \sum_j J_{ij}
\] 
and 
\[
J_{ij} = \frac{g_{ij}}{kT} (\mu_j - \mu_i).
\]
The external potential $V_j$ is set to zero in the main text.

It is interesting to note that
not the chemical potentials but the activities at the sites are reinterpreted as electric potentials, which leads to differences
in the behavior of the random resistor-capacitor model and the random potential model far from equilibrium, i.e. for large concentration gradients.  

\subsection{B - Decoupling of local concentrations and local activities under quasiequilibrium}

In a stationary situation corresponding to quasiequilibrium we have
\begin{equation}
\sum_j g_{ij}\zeta_j - \left( \sum_j g_{ij} \right) \zeta_i = 0, \label{means}
\end{equation}
where the sum runs over the nearest neighbors of the site $i$.
Moreover, in the thermodynamical limit $L \to \infty$ the permeability of the membrane tends to zero, 
and all local currents in it as well:
\begin{equation}
J_{ij} = g_{ij}(\zeta_j-\zeta_i) \to 0. \label{curr}
\end{equation}

Now we show that in quasi-equilibrium the local concentrations and the local activities (``potentials'') tend to be independent from
each other (although the local transition rates are correlated with the Boltzmann factors and thus local concentrations). 
To see this we note that according to Eq.(\ref{means}) the activity (potential) at a site $i$ is the weighted arithmetic mean
of the activities of the neighbors it is connected to, and Eq.(\ref{curr}) states that the difference between the
activities of the connected sites under quasi-equilibrium tends arbitrarily small. Therefore our system can be
considered as composed of large, practically equipotential regions whose potential hardly fluctuates
around its mean $\zeta^*$ depending on the region's position. In these regions the concentrations $n_i = \zeta_i n_i^0$
can be averaged over the physically small volume still containing the large number of sites, so that
$\langle n \rangle = \zeta^* \langle n_i^0 \rangle$. We note that this last
property does not rely on homogenization or on the isotropy and will hold even if our system is built of independent 
parallel or interwoven wires!

Let us now assume that at large scales the corresponding electric system 
homogenizes. In this case the total voltage profile (which in our case is mapped on the
profile of activities) obtained by the coarse graining 
(moving average) of the voltages (activities) over macroscopic domains of the system which are large enough 
compared to the lattice spacing but small compared to $L$ follows a linear behavior (similar to those of the
effective concentration in Fig.1).
Then the coarse-grained activity is a linear function of the coordinate, and thus also the coarse grained concentration
is linear in coordinate, with the 
proportionality factor $\langle n_i^0 \rangle$ between the both.

\subsection{C - Discretization of a continuous Fokker-Planck equation with random potential}

We start this section by showing the discretization procedure linking the Fokker-Planck equation 
to the master equation. The equation in a continuous $d$-dimensional space is
discretized using a regular hypercubic lattice with lattice constant $a$ considered sufficiently small. 
The sites are identified by the vectors $\mathbf{i}$, the site energies $E_{\mathbf{i}}$ are given by
corresponding values of the random potential $U(\mathbf{x})$ at the position of site $\mathbf{i}$, 
$\mathbf{x} = \mathbf{i} = (i_1, ...,i_d) = a(n_1,... ,n_d)$ where $n_k$ 
is an integer number and $k$ runs from 1 to $d$ .
Correspondingly, the derivatives $\partial_k$ along a certain direction $k$
are linked to the forward differences $\Delta_k$ in the following way:
\begin{eqnarray}
 \partial_k p(\mathbf{x}) &\rightarrow&  \frac{1}{a} \Delta_k p_{\mathbf{i}} 
=  \frac{1}{a} [p_{\mathbf{i} + a\hat{k}} - p_{\mathbf{i}}] \nonumber \\
 \partial_k^2 p(\mathbf{x}) &\rightarrow&  \frac{1}{a^2} \Delta^2_k p_{\mathbf{i}} 
=  \frac{1}{a^2} [p_{\mathbf{i} + a\hat{k}} - 2 p_{\mathbf{i}} + p_{\mathbf{i} - a\hat{k}}]
\end{eqnarray}
where $\hat{k}$ is the unit vector in the $k$-th direction.
Hence, from
\begin{eqnarray*}
\dot{p}(\mathbf{x}) &=& D \nabla^2 p(\mathbf{x}) + D \beta \nabla [\nabla U(\mathbf{x}) \cdot p(\mathbf{x})] = \\ 
&=& D \sum_{k=1}^d \Big[  \partial^2_k p(\mathbf{x}) + \beta \partial_k [\partial_k U(\mathbf{x}) \cdot p(\mathbf{x})]\Big] 
\end{eqnarray*}
follows 
\begin{eqnarray}
\label{1}
\dot{p}_{\mathbf{i}} &=& D \sum_{k=1}^d \Big[ \frac{1}{a^2} [p_{\mathbf{i} +a\hat{k}} - 2 p_{\mathbf{i}} + p_{\mathbf{i} -a\hat{k}}] + \nonumber \\
&&+ \frac{\beta}{a}\partial_k \Big( (E_{\mathbf{i} +a\hat{k}} - E_{\mathbf{i}}) p_{\mathbf{i}}\Big)\Big]   
\end{eqnarray}
We consider the potential to be a slowly varying function 
of the position; this allows to substitute the 
remaining partial derivative with half of a double step 
forward difference, the consequent error being 
of the smaller order of magnitude $ \Delta_k^2 E$. 
\begin{eqnarray}
\dot{p}_{\mathbf{i}} &=& \frac{D}{a^2} \sum_{k=1}^d \Big[  [p_{\mathbf{i} +a\hat{k}} - 2 p_{\mathbf{i}} + p_{\mathbf{i} -a\hat{k}}] +  \\
&& + \frac{\beta}{2}  \Big( (E_{\mathbf{i} + 2a\hat{k}} - E_{\mathbf{i} +a\hat{k}}) p_{\mathbf{i} + a\hat{k}} - 
(E_{\mathbf{i}} - E_{\mathbf{i}-a\hat{k}}) p_{\mathbf{i}-a\hat{k}}\Big)\Big]  \nonumber
\end{eqnarray}
Neglecting again second order terms we replace the quantity
$(E_{\mathbf{i} + 2a\hat{k}} - E_{\mathbf{i} +a\hat{k}})$ with 
$(E_{\mathbf{i} +a\hat{k}} - E_{\mathbf{i}})$, 
and add and subtract  the term
$(E_{\mathbf{i} +a\hat{k}} - E_{\mathbf{i}}) p_{\mathbf{i}}$
 \begin{eqnarray}
\dot{p}_{\mathbf{i}}  &=& \frac{D}{a^2} \sum_{k=1}^d \Big[  [p_{\mathbf{i} +a\hat{k}} - 2 p_{\mathbf{i}} + p_{\mathbf{i} -a\hat{k}}] + \nonumber \\
&&+ \frac{\beta}{2}  \Big( (E_{\mathbf{i} +a\hat{k}} - E_{\mathbf{i}}) p_{\mathbf{i}+a\hat{k}} - 
(E_{\mathbf{i}} - E_{\mathbf{i} -a\hat{k}}) p_{\mathbf{i} -a\hat{k}} + \nonumber \\
&& + (E_{\mathbf{i} +a\hat{k}} - E_{\mathbf{i}}) p_{\mathbf{i}} 
- (E_{\mathbf{i} +a\hat{k}} - E_{\mathbf{i}}) p_{\mathbf{i}}\Big)\Big] =\\
&=& \frac{D}{a^2} \sum_{k=1}^d \Big[  [p_{\mathbf{i} +a\hat{k}} - 2 p_{\mathbf{i}} + p_{\mathbf{i} -a\hat{k}}] + \nonumber \\
&&+ \frac{\beta}{2}  \Big( (E_{\mathbf{i} +a\hat{k}} - E_{\mathbf{i}}) p_{\mathbf{i}+a\hat{k}} - 
(E_{\mathbf{i}} - E_{\mathbf{i} -a\hat{k}}) p_{\mathbf{i} -a\hat{k}} + \nonumber \\
&& + (E_{\mathbf{i} +a\hat{k}} - E_{\mathbf{i}}) p_{\mathbf{i}} 
+ (E_{\mathbf{i} -a\hat{k}} - E_{\mathbf{i}}) p_{\mathbf{i}}\Big)\Big] =  \\
&=& \frac{D}{a^2} \sum_{k=1}^d \Big[  
[p_{\mathbf{i} +a\hat{k}}\Big(1 - \frac{\beta}{2}   (E_{\mathbf{i}} - E_{\mathbf{i} + a\hat{k}})\Big) +  \nonumber \\
&& + p_{\mathbf{i} -a\hat{k}}\Big(1 - \frac{\beta}{2}   (E_{\mathbf{i}} - E_{\mathbf{i} -a\hat{k}})\Big) +  \\
&& - p_{\mathbf{i}}\Big(2 - \frac{\beta}{2}   (E_{\mathbf{i} +a\hat{k}} - E_{\mathbf{i}}) 
 - \frac{\beta}{2}   (E_{\mathbf{i} -a\hat{k}} - E_{\mathbf{i}})\Big)\Big]  \nonumber
\end{eqnarray}
Since the difference between the energies at neighbouring sites is small, in the
lowest nonvanishing order we can put
\begin{eqnarray}
\dot{p}_{\mathbf{i}}  &=& \frac{D}{a^2} \sum_{k=1}^d \Big[  
[p_{\mathbf{i} +a\hat{k}} e^{- \frac{\beta}{2}(E_{\mathbf{i}} - E_{\mathbf{i} + a\hat{k}})} +  \nonumber \\
&+&  p_{\mathbf{i} -a\hat{k}}e^{ - \frac{\beta}{2}(E_{\mathbf{i}} - E_{\mathbf{i} - a\hat{k}})} + \nonumber  \\
&-& p_{\mathbf{i}} \Big( e^{ -\frac{\beta}{2}(E_{\mathbf{i} +a\hat{k}} - E_{\mathbf{i}})} 
+ e^{ - \frac{\beta}{2}(E_{\mathbf{i} -a\hat{k}} - E_{\mathbf{i}})} \Big)\Big] 
\end{eqnarray}
At this point, the discretized Fokker-Planck equation 
is shown to be equivalent to the master equation 
\begin{eqnarray}
\dot{p}_{\mathbf{i}} &=& \sum_{k=1}^d \Big[ w_{\mathbf{i},\mathbf{i} + a\hat{k}}p_{\mathbf{i} +a\hat{k}} +
w_{\mathbf{i},\mathbf{i} -a\hat{k}}p_{\mathbf{i} -a\hat{k}} + \nonumber \\
&-&  p_{\mathbf{i}}(w_{\mathbf{i} + a\hat{k},\mathbf{i}} + w_{\mathbf{i} - a\hat{k}, \mathbf{i}}) \Big]
\end{eqnarray}
by setting
$$
w_{\mathbf{i},\mathbf{j}} = \frac{D}{a^2} e^{ - \frac{\beta}{2}(E_{\mathbf{i}} - E_{\mathbf{j}})} 
 = w_0  e^{ - \frac{\beta}{2}(E_{\mathbf{i}} - E_{\mathbf{j}})}
$$
implying that, up to constant factors
$$
g_{\mathbf{i},\mathbf{j}} = w_{\mathbf{i},\mathbf{j}} e^{ - \beta E_{\mathbf{j} }}= w_0 e^{ - \frac{\beta}{2}(E_{\mathbf{i}} + E_{\mathbf{j}})}.
$$
In what follows $\frac{\beta}{2}(E_{\mathbf{i}} + E_{\mathbf{j}})$ can be changed for
$U(\mathbf{x})/kT$. Thus, returning to the continuous limit when calculating the macroscopic
(effective medium) conductance of the continuous disordered medium we can take its local
conductivity to be $g(\mathbf{x})= g_0 \exp(-U(\mathbf{x})/kT)$ with $g_0 = D/a^2$.

\subsection{D - Calculations pertinent to Figs. 2 and 3}

We now proceed to compute the Hashin-Shtrikman 
bounds for the effective diffusion constant by considering 
two different distributions of the energy variable $E_{\mathbf{i}}$.

\subsubsection{Binary distribution}

In the binary case
\begin{equation}
E_{\mathbf{i}}=
\left\{
\begin{array}{rl}
E_{1} & \mbox{ with probability} \quad p_1 = p \\
E_{2} & \mbox{ with probability} \quad p_2 = 1- p
\end{array}
\right.
\end{equation}
with $E_2 < E_1$. 
The local conductance follows again a binary distribution
\begin{equation}
g_{\mathbf{i},\mathbf{j}} =
\left\{
\begin{array}{rl}
g_1 \propto g_0 e^{-\beta E_1} & \mbox{ with probability} \quad p_1 = p \\
g_2 \propto g_0 e^{-\beta E_2} & \mbox{ with probability} \quad p_2 = 1-p
\end{array}
\right.
\end{equation}
or, equivalently,
\begin{equation}
p(g) = p_1 \delta(g - g_1) + p_2 \delta(g - g_2). 
\end{equation}
Here we introduce the following notation for different averages
which will repeatedly appear in what follows (weighted arithmetic, arithmetic and geometric mean):
\begin{eqnarray}
\langle g \rangle &=& p_1g_1 + p_2g_2 \nonumber \\
\langle g \rangle_A &=& \frac{1}{2}(g_1 + g_2) \nonumber \\
\langle g \rangle_G &=& \sqrt{g_1g_2} \nonumber \\
\end{eqnarray}
The notation $\langle f(E) \rangle$ will be continuously used for the mean of
any physical quantity $f$ over the energy distribution. 
The Effective Medium Approximation then gives the effective conductance $g^* $ 
as the solution for the self-consistency condition
\begin{equation}\label{scc}
\left\langle \frac{g^* - g}{(d-1)g^* + g} \right\rangle = 0
\end{equation}
resulting in
\begin{equation}\label{bingeff}
g^* = \frac{d\langle g \rangle - 2 \langle g \rangle_A }{2(d-1)}
\left[\; 1 + \sqrt{1 + \frac{4(d-1)\langle g \rangle_G^2}
{[d\langle g \rangle - 2 \langle g \rangle_A]^2}} \; \right]
\end{equation}
The percolating case is easily recovered 
in the limit 
\newline
$E_1 \rightarrow +\infty$, $g_1 \rightarrow 0$, 
giving the wellknown result ([14])
\begin{equation}
g^* = \frac{dp_2-1}{d-1}g_2 \qquad 
p_2^c = 1/d
\end{equation}
while simple calculations can show that in the two-dimensional
symmetric case, $d=2$, $p=1/2$, 
$g^*$ equals the geometric mean $\sqrt{g_1g_2}$, as stated in the main text.
\newline
Eq.(\ref{bingeff}) is finally used to calculate 
the effective diffusion constant
\begin{eqnarray}\label{Deff}
D^* &=& \frac{a^2 g^*}{\langle e^{-\beta E_{\mathbf{i}}} \rangle} =  \frac{D g^*}{\langle g \rangle}  = \\
&=& D \frac{d\langle g \rangle - 2 \langle g \rangle_A }{2(d-1)\langle g \rangle}
\left[\; 1 + \sqrt{1 + \frac{4(d-1)\langle g \rangle_G^2}
{[d\langle g \rangle - 2 \langle g \rangle_A]^2}} \; \right] \nonumber
\end{eqnarray}
Following the strategy outlined in [11], 
we now introduce a free parameter $g'$ and define the quantity 
\begin{equation}
B(g') = \left\langle \frac{g - g'}{(d-1)g' + g} \right\rangle. 
\end{equation}
According to [11], the following inequalities hold: 
\begin{eqnarray}
g^* &>& g' \Big( 1 + \frac{d B(g')}{1 - B(g')}\Big)
\quad \mbox{if} \quad g' < \min (g) = g_1 \nonumber \\ 
g^* &<& g' \Big( 1 + \frac{d B(g')}{1 - B(g')}\Big) 
\quad \mbox{if} \quad g' > \max (g) = g_2 \nonumber 
\end{eqnarray}
Denoting
\begin{eqnarray}
g_1^* &=& g_1 \Big( 1 + \frac{d B(g_1)}{1 - B(g_1)}\Big) \\
g_2^* &=& g_2 \Big( 1 + \frac{d B(g_2)}{1 - B(g_2)}\Big)
\end{eqnarray}
we get the following inequalities 
\begin{eqnarray}
g^* &>& g_1^* = g_1\Big[ 1 + \frac{d(1-p)(g_2 - g_1)}{dg_1 + p(g_2 - g_1)}\Big] \\
g^* &<& g_2^* = g_2\Big[ 1 - \frac{dp(g_2 - g_1)}{dg_2 -(1- p)(g_2 - g_1)}\Big] 
\end{eqnarray}
and define
\begin{eqnarray}
D_1^* &=& \frac{a^2 g_1^*}{\langle e^{-\beta E_{\mathbf{i}}} \rangle} = 
D \frac{ g_1}{\langle g \rangle}\Big[ 1 + \frac{d(1-p)(g_2 - g_1)}{dg_1 + p(g_2 - g_1)}\Big]  \\
D_2^* &=& \frac{a^2 g_2^*}{\langle e^{-\beta E_{\mathbf{i}}} \rangle} =
D \frac{ g_2}{\langle g \rangle}\Big[ 1 - \frac{dp(g_2 - g_1)}{dg_2 -(1- p)(g_2 - g_1)}\Big]  \nonumber
\end{eqnarray}
giving the bounds
\begin{equation}
D_1^* <  D^* < D_2^*  
\end{equation}
Through a simple rescaling of the units 
we can fix $D = g_1= 1$ and express 
$g_2$ via the contrast $x=g_2/g_1$ and $g^*$ via its rescaled value 
$\tilde{g}^*= g^*/g_1$, to get a better graphical representation.
The corresponding boundaries and the EMA result for $\tilde{D}^*$ then read  
\begin{eqnarray}\label{Deff2}
\tilde{D}^* &=& \frac{d[p + (1-p)x] - (x+1)}{2(d-1)[p + (1-p)x]} \times \\
& \times & \left[\; 1 + \sqrt{1 + \frac{4(d-1)x} 
{[d[p + (1-p)x] - (x+1)]^2}} \; \right]  \nonumber 
\end{eqnarray}
\begin{eqnarray}
\tilde{D}_1^* &=&  \frac{1}{p + (1-p)x} \Big[ 1 + \frac{d(1-p)(x - 1)}{d + p(x - 1)}\Big] \\ 
\tilde{D}_2^* &=& \frac{x}{p + (1-p)x}  \Big[ 1 - \frac{d p(x - 1)}{dx - (1- p)(x - 1)}\Big]  
\end{eqnarray}
with
\begin{equation}
\tilde{D}_1^* <  \tilde{D}^*  < \tilde{D}_2^* 
\end{equation}

These bounds, together with the effective diffusivity of eq.(\ref{Deff2}), 
are shown in figure 2 of the main text for two- and three-dimensional cases and for $p=1/2$.

\subsubsection{Truncated exponential distribution}

Let us now consider the case in which the possible values 
of the site energies $E_{\mathbf{i}}$ are exponentially 
distributed in the interval $[E_2, E_1]$ 
\begin{equation}
P(E_{\mathbf{i}}) = \frac{\beta \, e^{-\beta E_{\mathbf{i}}}}{e^{-\beta E_2} - e^{-\beta E_1}}.
\end{equation}
Performing the change of variables it is easy to show that the bond conductance is a random variable 
uniformly distributed in the interval $[g_1 = g_0e^{-\beta E_1}\,, \;  g_2 = g_0e^{-\beta E_2}]$.
Eq.(\ref{scc}) then gives the following condition
to be satisfied by the effective conductance
\begin{equation}
\Big[ (d-1)g^* + g_2\Big] e^{-g_2/dg^*} = \Big[ (d-1)g^* + g_1\Big] e^{-g_1/dg^*} 
\end{equation}
or by the effective diffusivity $D^*$                    
\begin{eqnarray}
\Big[ (d-1)\langle g \rangle D^*  + g_2 D\Big] e^{-g_2 D/(d\langle g \rangle D^*) } = && \nonumber \\
= \Big[ (d-1)\langle g \rangle D^* + g_1 D\Big] e^{-g_1D/(d\langle g \rangle D^*) } && 
\end{eqnarray}
which can be rewritten through the previous rescaling as
\begin{eqnarray}
\Big[ (d-1)(x+1)\tilde{D}^* + 2x \Big] e^{- 2x/(d(x+1)\tilde{D}^*)} = && \nonumber \\ 
= \Big[ (d-1)(x+1)\tilde{D}^* +2\Big] e^{-2/(d(x+1) \tilde{D}^*)} &&
\end{eqnarray}
\medskip
\newline
The solution $\tilde{D}^*$ of this equation is 
shown graphically in Fig. 3 of the main text  together 
with the two bounds we get ready to calculate.
\newline
We consider again 
\begin{equation}
B(g') = \left\langle \frac{g - g'}{(d-1)g' + g} \right\rangle
\end{equation}
and the inequalities
\begin{eqnarray}
g^* &>& g' \Big( 1 + \frac{d B(g')}{1 - B(g')}\Big)
\quad \mbox{if} \quad g' < \min (g_{ij}) = g_1 \nonumber \\ 
g^* &<& g' \Big( 1 + \frac{d B(g')}{1 - B(g')}\Big) 
\quad \mbox{if} \quad g' > \max (g_{ij})= g_2 \nonumber 
\end{eqnarray}
In this case we have
\begin{equation}
B(g') = 1 - \frac{dg'}{g_2 - g_1} 
\ln\Big[ \frac{(d-1)g' + g_2}{(d-1)g' + g_1}\Big]
\end{equation}
and setting respectively $g' = g_1$ and $g' = g_2$, 
we can write
\begin{eqnarray}
B(g_1) &=& 1 - \frac{dg_1}{g_2 - g_1}\ln\Big[ \frac{dg_1 + g_2 - g_1}{dg_1}\Big] \nonumber  \\
&=& 1 -  \frac{1}{y_1} \ln\Big(1 + y_1\Big)  \\
B(g_2) &=& 1 - \frac{dg_2}{g_2 - g_1}\ln\Big[ \frac{dg_2}{dg_2 - g_2 + g_1}\Big] \nonumber  \\
&=& 1 + \frac{1}{y_2}\ln\Big(1 - y_2\Big)
\end{eqnarray}
where the couple of parameters
\begin{equation}
y_1 = \frac{g_2 - g_1}{dg_1}  \qquad
y_2 = \frac{g_2 - g_1}{dg_2} 
\end{equation}
has been introduced. 
At the end the two bounds for the conductivity 
are given by
\begin{eqnarray}
g^* &>& g_1^* =  g_1\Big[ 1 + d \Big(\frac{y_1}{\ln(1 + y_1)} - 1\Big)\Big] \\
g^* &<&  g_2^* =  g_2\Big[ 1 - d \Big(\frac{y_2}{\ln(1 - y_2)} + 1\Big)\Big] 
\end{eqnarray}
from which we obtain
\begin{eqnarray}
D_1^* &=&    \frac{D}{\langle g \rangle}g_1\Big[ 1 + d \Big(\frac{y_1}{\ln(1 + y_1)} - 1\Big)\Big] \\
D_2^* &=&   \frac{D}{\langle g \rangle}g_2\Big[ 1 - d \Big(\frac{y_2}{\ln(1 - y_2)} + 1\Big)\Big]  
\end{eqnarray}
and the following bounds 
\begin{equation}
D_1^* <  D^*  <  D_2^* 
\end{equation}
Then, fixing again $D = g_1=1$, $g_2=x$, $g^* = \tilde{g}^*$, 
we write
\begin{eqnarray}
\tilde{D}_1^* &=& \frac{2}{x+1}  \Big[ 1 - d + \frac{x - 1}{\ln(1 + \frac{x - 1}{d})} \Big]  \\
\tilde{D}_2^* &=&   \frac{2x}{x+1}   \Big[ 1 - d  - \frac{x - 1}{x \ln(1 - \frac{x - 1}{dx})} \Big] 
\end{eqnarray}
with
\begin{equation}
\tilde{D}_1^* <  \tilde{D}^*  < \tilde{D}_2^* 
\end{equation}

\end{document}